\documentclass{PoS}
\usepackage{epsfig,dsfont,nicefrac}           

\title{Remarks on left-handed lattice fermions
\thanks{Based on presentations by both authors.}}

\ShortTitle{Left-handed lattice fermions}

\author{Christof Gattringer\\
        University of Graz, Austria\\
        E-mail: \email{christof.gattringer@uni-graz.at}}

\author{Markus Pak\\
        Universitiy of Graz, Austria\\
        E-mail: \email{markus.pak@stud.uni-graz.at}}

\abstract{We study whether applying lattice projectors on a vector-like 
Ginsparg-Wilson Dirac operator is the only way to construct left-handed
lattice fermions. Using RG transformations we derive an equation for the 
generating functional on the lattice, obtained by blocking from the 
continuum. We analyze how symmetries of the continuum
theory manifest themselves in this lattice generating functional and how
anomalies emerge. The formalism is applied to left-handed continuum 
fermions and we derive two equations that characterize the
corresponding lattice symmetries. To analyze possible solutions, we 
directly block a parameterized action for free continuum fermions to obtain the 
corresponding lattice action. Subsequently we study parameter 
values where the continuum action corresponds to left-handed fermions.}

\FullConference{The XXV International Symposium on Lattice Field Theory\\
                 July 30 - August 4, 2007\\
                 Regensburg, Germany}

\begin{document}

\section{Introduction}
Understanding the implementation of chiral symmetry for vector-like
theories on the lattice was one of the major achievements in this
field. The central equation which governs chiral symmetry on the
lattice is the Ginsparg-Wilson (GW) relation \cite{GiWi82} which a
lattice Dirac operator $D$ has to obey
\begin{equation}
\gamma_5 D \; + \; D \, \gamma_5 \; = \; D \gamma_5 D \; .
\label{giwi}
\end{equation}
This quadratic equation is the lattice manifestation of the
anti-commutator $\gamma_5 {\cal D} + {\cal D} \gamma_5 = 0$, which a
massless vector-like Dirac operator ${\cal D}$ obeys in the
continuum. Based on this equation a wealth of beautiful structure
was discovered, such as the lattice version of chiral symmetry
\cite{luscher} and the relation to gauge field topology through the
index theorem and the axial anomaly \cite{hasenfratz1}. 
Solutions of (\ref{giwi}) are given by the overlap operator
\cite{overlap} and fixed-point fermions \cite{fp}.

Once the vector-like chiral symmetry was understood on the lattice,
this opened the door towards the construction of chiral, e.g.,
left-handed lattice fermions. The goal of such a construction is to
find the lattice form of left-handed fermions,
described by a projected continuum Dirac operator
\begin{equation}
{\cal D}_- \; = \; \frac{ \mathds{1} + \gamma_5}{2}\; {\cal D} \; =
\; {\cal D} \, \frac{ \mathds{1} - \gamma_5}{2} \; .
\label{contleft}
\end{equation}

A natural candidate for a left-handed lattice Dirac operator is
obtained by projecting a vector-like solution $D$ of the
Ginsparg-Wilson equation:
\begin{equation}
D^a_- \; = \; \frac{ \mathds{1} +  \gamma_5 }{2} \, D \; = \; D \,
\frac{\mathds{1} -  \gamma_5 [ \mathds{1} - D ] }{2} \; .
\label{old}
\end{equation}
A projector $[\mathds{1} +  \gamma_5]/2$ is applied from the left.
When using the Ginsparg-Wilson equation (\ref{giwi}),
the projection from the left can be
rewritten into a left-handed projector with a modified 
$\gamma_5$-term, 
acting from the right. This type of projection and its
consequences for chiral gauge theories were analyzed in various
papers (see \cite{chiralgauge} for a selection).

Since in (\ref{old}) the projections from the left and from the right
appear in an asymmetric way, we 
refer to the construction (\ref{old}) as the {\it
asymmetrically projected Dirac operator} and indicate the asymmetry
by the superscript $a$ in $D^a_-$.

An interesting aspect is the fact that the projected operator
$D^a_-$ does not obey the GW equation (\ref{giwi}). This is
potentially worrisome since the left-handed continuum operator
${\cal D}_-$ does anti-commute with $\gamma_5$,
\begin{equation}
\gamma_5 \, {\cal D}_- \; + \; {\cal D}_- \, \gamma_5 \; = \; 0 \; ,
\label{contcommut}
\end{equation}
 and the GW
equation is the lattice manifestation of this anti-commutator. This
observation raises the question whether all of the symmetry of
left-handed continuum fermions has been transported onto the lattice
in an optimal way. A possible source of complications might be the
fact that first the symmetries of a vector-like theory were
transported onto the lattice and the chiral projectors were then
constructed from the vector-like objects.

In this paper we study whether first constructing the vector-like lattice
Dirac operator which obeys (\ref{giwi}), followed by the subsequent projection 
(\ref{old}), is the only way to obtain left-handed lattice fermions. 
As an alternative idea we omit the intermediate step 
of a vector-like lattice theory and analyze 
the possibility to directly map the symmetries of left-handed continuum
fermions onto the lattice. We show that left-handed lattice
fermions should obey two equations which have a structure similar to
the Ginsparg-Wilson relation, i.e., are quadratic equations for a
left-handed lattice Dirac operator $D_-$. We refer to these
equations as the {\it chiral Ginsparg-Wilson equations} ($\chi$-GW).
We demonstrate that if a left-handed lattice Dirac operator
$D_-$ obeys both of them, it also obeys the vector-like GW equation,
i.e., implements correctly also the anti-commutator
(\ref{contcommut}) from the continuum. Assuming the existence of a solution 
of the $\chi$-GW equations, we show that a closed algebra of projectors
emerges. 

We analyze the relation between the continuum symmetries and their lattice
counterparts further by blocking free continuum fermions onto the
lattice. The continuum action we use, as well as our blocking kernel contain free
parameters. In the action these parameters allow to interpolate between 
vector-like and chiral fermions, while in the blocking kernel the parameters can be
used to switch on and off the mixing of different components in the blocking
procedure. The lattice fermion action is obtained as a function of our
parameters and we discuss different limits of their values which are related
to left-handed fermions.
 
\section{Mapping continuum symmetries onto the lattice}

In this section we discuss how symmetries of the continuum theory manifest
themselves in the generating functional on the lattice, which is constructed
by blocking from the continuum. 

\subsection{The generating functional and its symmetries}
The starting point of our analysis is the equation which defines the
Dirac operator $D$ on the lattice through a blocking transformation
from the fermionic continuum action $S_F$:
\begin{equation}
e^{- \,\overline{\psi} D \psi} \; = \; \int D[ \overline{\Phi},
\Phi] \, e^{-\, S_F[ \overline{\Phi},\Phi] \, - \, [\overline{\psi}
- \overline{\Phi}^B] \, B \, [ \psi - \Phi^B ]} \; . \label{effact}
\end{equation}
The lattice fermions are denoted by $\overline{\psi}$, $\psi$ and we
use vector/matrix notation for all indices (space-time, color,
Dirac). $\overline{\Phi}, \Phi$ are the fermion fields in the
continuum which enter the path integral on the right-hand side.
Through integrating over hypercubes which are centered at the points
of the lattice one constructs from them the blocked fields
$\overline{\Phi}^B, \Phi^B$. These then live on the sites of the
lattice and consequently they have the same indices as the lattice
fields $\overline{\psi}, \psi$, in particular a discrete space-time
index. The blocking kernel $B$ determines how the blocked fields
$\overline{\Phi}^B, \Phi^B$ and the lattice fields $\overline{\psi},
\psi$ are mixed in the bilinear form in the exponent. We remark,
that Eq.~(\ref{effact}) is understood in a background gauge field
which also has to be blocked in a suitable way \cite{wiese}.
The work by Ginsparg and Wilson \cite{GiWi82} starts from
Eq.~(\ref{effact}) and analyzes its behavior under a chiral rotation
of the lattice fields. Hasenfratz et al explore this equation by
analyzing its saddle point \cite{hasenfratz}.

Here we take a slightly different approach and consider the
generating functional on the lattice defined as
\begin{equation}
W[ \overline{J},J] \; = \; \int  D[\overline{\psi}, \psi] \,
e^{-\overline{\psi} D \psi \, + \, \overline{\psi} J \, + \,
\overline{J} \psi} \; , \label{generfunct1}
\end{equation}
where we have coupled sources $\overline{J}$ and $J$ to the lattice
fermions. Inserting the exponential of the action from Eq.\
(\ref{effact}) we find an expression for the generating functional
through a blocking prescription:
\begin{eqnarray}
W[\overline{J},J] &  =  & \int\!\!D[\overline{\psi}, \psi]
\, e^{\overline{\psi} J \, + \, \overline{J} \psi} \,
\int\!\!D[\overline{\Phi}, \Phi] \, e^{- \,
S_F[\overline{\Phi},\Phi] \, - \, [\overline{\psi} -
\overline{\Phi}^B] \, B \, [ \psi - \Phi^B ] }
\nonumber \\
&  =  & \int\!\!\!D[\overline{\Phi}, \Phi] \, e^{-
\overline{\Phi}^B\!  B  \Phi^B
 -  S_F[\overline{\Phi},\Phi]} \!
\int\!\!  D[\overline{\psi}, \psi] \, e^{-  \overline{\psi} B \psi
\, + \, \overline{\psi} [ J + B\Phi^B]  \,+\, [\overline{J} +
\overline{\Phi}^B\! B ] \psi}
\nonumber \\
&  =  & \det[B] \, e^{\overline{J} B^{-1} J} \int\!\!
D[\overline{\Phi}, \Phi] \, e^{\overline{J} \Phi^B \, + \,
\overline{\Phi}^B J \, - \, S_F[\overline{\Phi},\Phi]} \; .
\label{generfunct2}
\end{eqnarray}
In the last step we have already solved the Gaussian integral over
the lattice fields and obtain an expression for the lattice
generating functional in terms of a continuum path integral.

We now explore how a symmetry of the continuum fermion action $S_F$
affects the lattice generating functional $W[\overline{J},J]$. In
particular we consider a transformation of the continuum fields,
\begin{equation}
\overline{\Phi} \, \rightarrow  \, \overline{\Phi}^\prime \, = \,
\overline{\Phi} \,  e^{i \varepsilon \overline{T}} \; \; \; \; , \;
\; \; \; \Phi \, \rightarrow  \, \Phi^\prime \, = \, e^{i
\varepsilon T} \Phi \; . \label{conttrafo}
\end{equation}
The generators $\overline{T}, T$ of the transformations we consider
here are Dirac matrices\footnote{More general transformations are
possible (see
  \cite{hasenfratz}).}.
This implies that the blocked fields $\overline{\Phi}^B$ and
$\Phi^B$ transform in the same way, since the blocking from
$\overline{\Phi}, \Phi$ to $\overline{\Phi}^B, \Phi^B$ is a purely
scalar operation, in other words, the blocked fields are essentially
linear combinations of the continuum fields. We stress that the
generators $\overline{T}$ and $T$ are independent transformations,
i.e., $\overline{\Phi}$ and $\Phi$ need not transform in the same
way.

Let us now assume that the transformation (\ref{conttrafo}) is a
symmetry of the action,
\begin{equation}
S_F[\overline{\Phi}^\prime, \Phi^\prime] \; = \;
S_F[\overline{\Phi}, \Phi] \; . \label{actsymm}
\end{equation}
We can also evaluate the integral over the continuum fields in the last
line of (\ref{generfunct2}) using the transformed variables
$\overline{\Phi}^\prime, \Phi^\prime$. Doing so and exploring the
invariance of the action, we obtain
\begin{eqnarray}
&& \!\!\!\! \int\!\! D[\overline{\Phi}^\prime, \Phi^\prime] \, e^{-
\, S_F[\overline{\Phi}^\prime,\Phi^\prime] \, + \, \overline{J}
\Phi^{B\,\prime} \, + \, \overline{\Phi}^{B\, \prime} J }
\label{contintsymm} \\
= && \!\!\!\!  \int \!\! D[\overline{\Phi} e^{i \varepsilon
\overline{T}}, e^{i \varepsilon T} \Phi] \, e^{- \,
S_F[\overline{\Phi},\Phi] \, + \, \overline{J}  e^{i \varepsilon T}
\Phi^{B} \, + \, \overline{\Phi}^{B} e^{i \varepsilon \overline{T}}
J }
\nonumber \\
= && \!\!\!\! \Big[ \,1 + i \varepsilon {\cal A}_{\overline{T}T} +
O(\varepsilon^2)\,\Big] \, \int \!\! D[\overline{\Phi}, \Phi] \,
e^{- \, S_F[\overline{\Phi},\Phi] \, + \, \overline{J}  e^{i
\varepsilon T} \Phi^{B} \, + \, \overline{\Phi}^{B} e^{i \varepsilon
\overline{T}} J } \; . \nonumber
\end{eqnarray}
In the last step we have transformed the measure of the continuum
path integral and taken into account that the transformation
(\ref{conttrafo}) could be anomalous with the anomaly ${\cal
A}_{\overline{T}T}$ showing up in the Jacobian of the transformation
\cite{Fujikawa}. Since later we will evaluate all expressions up to
$O(\varepsilon)$ we have kept only the leading term of the Jacobian.
For non-anomalous transformations $\overline{T},T$ one has ${\cal
A}_{\overline{T}T} = 0$.

Inserting the result (\ref{contintsymm}) back into the expression
(\ref{generfunct2}) for the generating functional we find that the
continuum symmetry (\ref{conttrafo}), (\ref{actsymm}) implies the
following symmetry of the generating functional on the lattice:
\begin{equation}
W[\overline{J},J] \; = \; e^{\overline{J} [ B^{-1} - e^{i
\varepsilon T} B^{-1} e^{i \varepsilon \overline{T}}] J} \, \Big[
\,1 + i \varepsilon {\cal A}_{\overline{T}T} +
O(\varepsilon^2)\,\Big] \, W[\overline{J}e^{i \varepsilon T},e^{i
\varepsilon \overline{T}}J] \, . \label{genfunctsymm1}
\end{equation}
Equation (\ref{genfunctsymm1}) summarizes how a continuum symmetry
reflects itself in the lattice generating functional
$W[\overline{J},J]$ constructed through blocking.

\subsection{Identification of the corresponding lattice symmetries}

Having analyzed the manifestation of a continuum symmetry, we now
want to identify a corresponding symmetry on the lattice which also
fulfills the symmetry condition (\ref{genfunctsymm1}). For that
purpose we consider transformed lattice fields
\begin{equation}
\overline{\psi}^\prime \, = \, \overline{\psi} e^{i \varepsilon
\overline{M}} \; \; \; \; , \; \; \; \; \psi^\prime \, = \, e^{i
\varepsilon M} \psi \; . \label{latttrafo}
\end{equation}
The two transformations $\overline{M}, M$ are not yet known and we
want to identify how they depend on $\overline{T}, T$ and the
blocking kernel $B$. The fact that the transformation should be a
symmetry of the lattice action implies
\begin{equation}
\overline{\psi}^\prime \, D \, \psi^\prime \; = \; \overline{\psi}
e^{i \varepsilon \overline{M}} \, D \, e^{i \varepsilon M} \psi \;
\stackrel{!}{=} \; \overline{\psi} \, D \, \psi \; .
\label{latinvariance}
\end{equation}
For later use we remark that when expanded in $\varepsilon$, the
invariance condition (\ref{latinvariance}) at $O(\varepsilon)$
implies the commutation relation
\begin{equation}
\overline{M} \, D \; + \; D\,M \; = \; 0 \; . \label{latcommut}
\end{equation}

As we have done for the continuum expression in the last section, we
now express the generating functional (\ref{generfunct1}) in terms
of the transformed fields $\overline{\psi}^\prime, \psi^\prime$ and
explore the implications of the symmetry (\ref{latinvariance}),
\begin{eqnarray}
W[\overline{J},J] & = & \int  D[\overline{\psi}^\prime, \psi^\prime]
\, e^{-\, \overline{\psi}^\prime D \psi^\prime \, + \,
\overline{\psi}^\prime J \, + \, \overline{J} \psi^\prime} \; ,
\label{genfunctderiv} \\
& = & \det\!\Big[ e^{i \varepsilon \overline{M}}\Big]
\det\!\Big[e^{i \varepsilon M}\Big] \int  D[\overline{\psi}, \psi]
e^{- \, \overline{\psi} D \psi \, + \, \overline{\psi} e^{i
\varepsilon \overline{M}}  J \, + \, \overline{J} e^{i \varepsilon
M} \psi } \; . \nonumber
\end{eqnarray}
The two Jacobi determinants up front come from the transformation of
the measure on the lattice. Using the formula $\det A =
\exp(\mbox{tr} \log A )$ we can expand them as
\begin{equation}
 \det\!\Big[ e^{i \varepsilon \overline{M}}\Big]
\det\!\Big[e^{i \varepsilon M}\Big] \; = \; 1 \, + \, i \varepsilon
\, \mbox{tr} \, [ \, \overline{M} + M \,] \, + \, O(\varepsilon^2)
\; .
\end{equation}
Combining the last two equations we find the symmetry relation for
the generating functional which is implied by the lattice symmetry
(\ref{latinvariance}),
\begin{equation}
W[\overline{J},J] \; = \; \Big[\, 1 \, + \, i \varepsilon \,
\mbox{tr} \, [ \, \overline{M} + M \,] \, + \, O(\varepsilon^2) \,
\Big] \, W[\overline{J} e^{i \varepsilon M} ,e^{i \varepsilon
\overline{M}} J] \; . \label{genfunctsymm2}
\end{equation}
Now the key idea is to compare the symmetry condition
(\ref{genfunctsymm2}) from the lattice transformation to the
corresponding continuum relation (\ref{genfunctsymm1}). By setting
the two equal we obtain the equation
\begin{eqnarray}
e^{\overline{J} [ B^{-1} - e^{i \varepsilon T} B^{-1} e^{i
\varepsilon \overline{T}}] J} \, \Big[ \,1 + i \varepsilon {\cal
A}_{\overline{T}T} + O(\varepsilon^2)\,\Big] \, W[\overline{J}e^{i
\varepsilon T},e^{i \varepsilon \overline{T}}J] \qquad \qquad \quad
\label{symmatch1}
\\
= \; \Big[\, 1 \, + \, i \varepsilon \, \mbox{tr} \, [ \,
\overline{M} + M \,] \, + \, O(\varepsilon^2) \, \Big] \,
W[\overline{J} e^{i \varepsilon M} ,e^{i \varepsilon \overline{M}}
J] \; , \nonumber
\end{eqnarray}
which we can use to identify the lattice transformation
$\overline{M},M$ which matches the continuum transformation
$\overline{T},T$. The last step is to insert the explicit form of
the generating functional,
\begin{equation}
W[\overline{J},J] \; = \; \det[D]\, e^{\; \overline{J} D^{-1} J } \;
, \label{genexp}
\end{equation}
which is obtained by directly solving the Gaussian integral
(\ref{generfunct1}). When inserting  (\ref{genexp}), Equation
(\ref{symmatch1}) becomes (we drop the factor $\det[D]$ on both
sides)
\begin{eqnarray}
e^{ \; \overline{J} [ B^{-1} - e^{i \varepsilon T} B^{-1} e^{i
\varepsilon \overline{T}}] J} \, \Big[ \,1 + i \varepsilon {\cal
A}_{\overline{T}T } + O(\varepsilon^2)\,\Big] \, e^{ \;
\overline{J}e^{i \varepsilon T} \, D^{-1} \, e^{i \varepsilon
\overline{T}}J  } \qquad \qquad \quad \label{symmatch2}
\\
= \; \Big[\, 1 \, + \, i \varepsilon \, \mbox{tr} \, [ \,
\overline{M} + M \,] \, + \, O(\varepsilon^2) \, \Big] \, e^{\;
\overline{J} e^{i \varepsilon M} \, D^{-1} \, e^{i \varepsilon
\overline{M}} J } \; . \nonumber
\end{eqnarray}
The last equation holds for arbitrary $\varepsilon$ and arbitrary
sources $\overline{J}, J$. Thus the terms bilinear in $\overline{J}$
and $J$, as well as the $O(\varepsilon)$ terms independent of
$\overline{J}, J$ have to match. For the latter term we conclude
\begin{equation}
{\cal A}_{\overline{T}T} \; = \; \mbox{tr} \, [ \, \overline{M} + M
\,] \; , \label{anomaly1}
\end{equation}
and thus have identified the lattice counterpart of the continuum
anomaly.

The terms bilinear in the sources $\overline{J}, J$ lead to a
symmetry relation for the quark propagator,
\begin{equation}
B^{-1} - e^{i \varepsilon T} B^{-1} e^{i \varepsilon \overline{T}}
\, + \, e^{i \varepsilon T} \, D^{-1} \, e^{i \varepsilon
\overline{T}} \; = \; e^{i \varepsilon M} \, D^{-1} \, e^{i
\varepsilon \overline{M}} \; .
\end{equation}
When expanding in $\varepsilon$ one obtains at $O(\varepsilon)$
\begin{equation}
T \, [ \, D^{-1} - B^{-1} \, ] \; + \; [ \, D^{-1} - B^{-1} \, ] \,
\overline{T} \; = \; M D^{-1} \, + \, D^{-1} \overline{M} \; .
\end{equation}
This equation is solved by
\begin{equation}
M \, = \, T \, [ \, \mathds{1} - B^{-1} D \, ] \qquad , \qquad
\overline{M} \, = \, [ \, \mathds{1} - D B^{-1} \, ] \, \overline{T}
\; . \label{mbarm}
\end{equation}
These are the generators of the lattice symmetry which we wanted to
construct. They depend on the continuum generators $\overline{T},T$,
the blocking kernel $B$ and the lattice Dirac operator $D$.
Inserting $\overline{M},M$ into the symmetry relation
(\ref{latcommut}) one ends up with a non-linear equation for the
lattice Dirac operator $D$,
\begin{equation}
\overline{T} D \, + \, D T \; = \; D\, [ \, B^{-1} \, \overline{T} +
T B^{-1} \, ] \, D \; , \label{gwgener}
\end{equation}
which is a generalization of the GW equation (see also
\cite{hasenfratz} for a similar result).

When inserting the explicit form (\ref{mbarm}) into the anomaly
equation (\ref{anomaly1}) one obtains the final form for the anomaly
\begin{equation}
{\cal A}_{\overline{T}T} \; = \; \mbox{tr} \, [ \, T + \overline{T}
-  TB^{-1}D - D B^{-1} \overline{T} \, ] \; . \label{anomaly2}
\end{equation}

It is easy to check that when blocking a vector-like continuum
theory with a blocking kernel $B = 2\cdot \mathds{1}$, and
considering a chiral rotation in the continuum, the equation
(\ref{gwgener}) reduces to the usual GW relation (\ref{giwi}), the
generators $\overline{M},M$ of (\ref{mbarm}) are the generators of
L\"uscher's symmetry \cite{luscher} and the anomaly assumes the form
\begin{equation}
{\cal A} \; = \; - \, \mbox{tr} \; \big[ \, \gamma_5 \, D \, \big] \; .
\end{equation}

\newpage
\section{Symmetries of left-handed fermions}

Having established the general connection between symmetries in the
continuum and their lattice counterparts, we can start to focus on
the case of left-handed fermions.

\subsection{Symmetries in the continuum}

To identify a
suitable continuum symmetry which is specific for left-handed fermions
we consider the Euclidean action
\begin{equation}
S[\overline{\Phi},\Phi] \; = \; \int d^4x \, \overline{\Phi}(x) \,
{\cal D}_- \, \Phi(x) \; . \label{contact1}
\end{equation}
The left-handed Dirac operator ${\cal D}_-$ in the continuum is
given by
\begin{equation}
{\cal D}_- \; = \; \gamma_\mu {\cal D}_\mu \, \frac{ \mathds{1} -
\gamma_5}{2} \; , \label{contact2}
\end{equation}
where ${\cal D}_\mu$ denotes the covariant derivative.

The action defined through (\ref{contact1}) and (\ref{contact2}) has
several symmetries. Here we focus on two symmetry transformations
which are not shared by a vector-like theory and are particular for
the left-handed action. We consider the two relations
\begin{eqnarray}
\mathds{1} \, {\cal D}_-  \; + \; {\cal D}_- \, \gamma_5  &  = & 0
\; , \label{symm1}
\\
\gamma_5 \, {\cal D}_- \; - \; {\cal D}_- \, \mathds{1} &  = & 0 \;
, \label{symm2}
\end{eqnarray}
which we will use for characterizing the left-handed Dirac operator
on the lattice. Summing (\ref{symm1}) and (\ref{symm2}) one obtains
the anti-commutator (\ref{contcommut}). We stress however, that the
anti-commutator (\ref{contcommut}) is not specific for left-handed
fermions, since it is also obeyed by the massless vector-like Dirac
operator. On the other hand  (\ref{symm1}) and (\ref{symm2}) are not
symmetries of the vector-like operator and thus are suitable for
characterizing left-handed fermions.

\subsection{The lattice counterparts of the continuum symmetries}

The next step is to identify the lattice symmetries that correspond
to (\ref{symm1}) and  (\ref{symm2}). We begin with the first
equation (\ref{symm1}). The corresponding symmetry transformation is
characterized by $\overline{T} = \mathds{1}$ and $T = \gamma_5$.
Using these in the expression (\ref{mbarm}), which identifies the
corresponding symmetry transformation on the lattice we obtain
\begin{equation}
M \, = \gamma_5 \, [ \, \mathds{1} - B^{-1} D_- \, ] \qquad , \qquad
\overline{M} \, = \, [ \, \mathds{1} - D_- B^{-1} \, ]  \; .
\label{mbarm1}
\end{equation}
In this formula $D_-$ denotes the Dirac operator of the target
theory, i.e., left-handed fermions on the lattice.

Inserting this into Eq.~(\ref{latcommut}), which expresses the
symmetry on the lattice, we obtain the following equation for the
Dirac operator:
\begin{equation}
\mathds{1} \, D_-  \; + \; D_- \, \gamma_5  \; = \; D_- \, [ \,
\mathds{1} \, + \, \gamma_5  \, ] \, B^{-1} \, D_- \; .
\label{latsymm1}
\end{equation}
This relation is the lattice equivalent of the continuum equation
(\ref{symm1}) and we refer to it (together with (\ref{latsymm2})
below) as the {\it chiral Ginsparg-Wilson relation} ($\chi$-GW).
Similar to the vector-like GW relation (\ref{giwi}), we have
obtained a quadratic piece on the right-hand side. The blocking
kernel $B$ has dimension of 1/length, such that when introducing the
lattice constant $a$ as scale, the right-hand side obtains an extra
factor of $a$. Thus in a naive continuum limit, $a \rightarrow 0$,
the right-hand side vanishes and the commutation relation
(\ref{symm1}) of the continuum is recovered.

In exactly the same way we can transport the second basic continuum
symmetry (\ref{symm2}) onto the lattice. In this case the continuum
generators are $\overline{T} = -\gamma_5$ and $T = \mathds{1}$ and
one obtains the corresponding lattice symmetry generators
\begin{equation}
\widetilde{M} \, = \, [ \, \mathds{1} - B^{-1} \widetilde{D}_- \, ]
\qquad , \qquad \widetilde{\overline{M}} \, = \, - \, \gamma_5 \, [
\, \mathds{1} - \widetilde{D}_- B^{-1} \, ]  \; . \label{mbarm2}
\end{equation}
We stress the fact that this symmetry corresponds to the second
continuum relation (\ref{symm2}) by denoting the corresponding
lattice symmetry generators by
$\widetilde{\overline{M}},\widetilde{M}$ and the lattice Dirac
operator by $\widetilde{D}_-$. The equation $\widetilde{D}_-$ has to
obey again follows from (\ref{latcommut}) and reads
\begin{equation}
\widetilde{D}_- \, \mathds{1}  \; - \; \gamma_5 \, \widetilde{D}_-
\; = \; \widetilde{D}_- \, B^{-1} \, [ \, \mathds{1} \, - \,
\gamma_5  \, ] \, \widetilde{D}_- \; . \label{latsymm2}
\end{equation}
This is the second $\chi$-GW relation.

The two  $\chi$-GW relations (\ref{latsymm1}) and (\ref{latsymm2})
are the lattice representations of the two symmetry transformations
(\ref{symm1}), (\ref{symm2}) which characterize left-handed
continuum fermions. It is important to note, that in the continuum
both (\ref{symm1}) and (\ref{symm2}) are obeyed by the left-handed
operator ${\cal D}_-$, while at this point it is not clear whether
on the lattice a simultaneous solution $D_- = \widetilde{D}_-$ of
both $\chi$-GW equations (\ref{latsymm1}), (\ref{latsymm2}) can be
found. We stress that the asymmetrically projected operator $D^a_-$
obeys only one of them, Eq.~(\ref{latsymm2}).

If for the moment we assume that such a common solution $D_- =
\widetilde{D}_-$ exists, then we can subtract (\ref{latsymm2}) from
(\ref{latsymm1}) and obtain
\begin{equation}
\gamma_5 \, D_- \; + \; D_- \, \gamma_5 \; = \; D_- \, [ \, B^{-1}
\, \gamma_5 \, + \, \gamma_5 \, B^{-1} \, ] \, D_- \; .
\label{chirgw}
\end{equation}
Thus if $D_-$ obeys both $\chi$-GW relations (\ref{latsymm1}),
(\ref{latsymm2}), it obeys also the vector-like GW equation
(\ref{chirgw}) which is the lattice form of the continuum
anti-commutator (\ref{contcommut}).

At this point we stress that the last argument works both ways: If a
lattice Dirac operator $D_-$ solves one of the two $\chi$-GW
relations and the vector-like GW (\ref{chirgw}), then it also solves
the other $\chi$-GW. Thus obeying any two of the equations
(\ref{latsymm1}), (\ref{latsymm2}) and (\ref{chirgw}) is equivalent
to obeying all three of them.

Let us conclude this section with a few remarks on possible solutions of the 
two $\chi$-GW relations (\ref{latsymm1}) and (\ref{latsymm2}). It is obvious,
that a Dirac operator of the form 
\begin{equation}
D_- \; = \; \frac{\mathds{1} + \gamma_5}{2} \, D_0 \,
\frac{\mathds{1} - \gamma_5}{2} \; ,
\end{equation}
which is obtained by projecting some lattice operator 
$D_0$ on both sides, trivially obeys (\ref{latsymm1}) and (\ref{latsymm2}),
since for both equations left- and right-hand sides vanish identically.  
Unfortunately such a double sided projection rules out a term proportional 
to $\mathds{1}$ which is needed for removing the doublers\footnote{We thank
Peter Hasenfratz for an interesting discussion on this point.}. At the
moment the structure of a joint solution for both $\chi$-GW equations, 
whether such a joint solution is possible at all, and how it is related to the
Nielsen-Ninomya result \cite{NiNo} is unclear. We stress however, that all results in the
section are for arbitrary blocking kernel $B$. This allows for several
interesting choices which still have to be explored.

\subsection{Chiral projectors}

We now show, that the approach of transferring the continuous
left-handed symmetries (\ref{symm1}), (\ref{symm2}) onto the lattice, 
gives rise to a consistent set of left- and right-handed lattice projectors. We
demonstrate this here for the projectors acting from the left, which
are constructed using (\ref{latsymm1}). Projectors acting from the
right can be obtained in the same way from (\ref{latsymm2}) and in
the end we quote the corresponding expressions.

It is obvious that (\ref{latsymm1}) can be rewritten in the form (a
factor of $1/2$ was multiplied)
\begin{equation}
D_- \, P_+ \; =  \; 0 \; , \label{dpright}
\end{equation}
with
\begin{equation}
P_+ \; = \; \frac{\mathds{1} + \gamma_5}{2} \, - \, \frac{\mathds{1}
+ \gamma_5}{2} \, B^{-1} \, D_- \; . \label{latpplus}
\end{equation}
Thus we can identify $P_+$ as a candidate for a right-handed
projector acting from the right.

A left-handed projector $P_-$, acting also from the right, which
corresponds to $P_+$ can be uniquely defined through $P_- =
\mathds{1} - P_+$. Explicitly $P_-$ is given by
\begin{equation}
P_- \; = \; \frac{\mathds{1} - \gamma_5}{2} \, + \, \frac{\mathds{1}
+ \gamma_5}{2} \, B^{-1} \, D_- \; . \label{latpminus}
\end{equation}

In order to establish that $P_+$ and $P_-$ are proper projectors one
needs to show that they obey
\begin{equation}
P_+ \, + \, P_- \, = \, \mathds{1} \quad , \quad P_\pm^2 \, = \,
P_\pm \quad , \quad P_+ \, P_- \, = \, P_- \, P_+ \, = \, 0 \; .
\label{projprop1}
\end{equation}
The first of these relations is trivially obeyed due to the
definition of $P_- = \mathds{1} - P_+$. The other properties are a
direct consequence of the $\chi$-GW relation (\ref{latsymm1}): One
evaluates the product of two projectors and the emerging piece
quadratic in $D_-$ is then rewritten using (\ref{latsymm1}). The
properties listed in (\ref{projprop1}) follow.

For constructing the projectors acting on the left, which one
obtains from $\widetilde{D}_-$, one proceeds in exactly the same way
(exploring Eq.\ (\ref{latsymm2})), and we only quote the final
expressions:
\begin{eqnarray}
&& \widetilde{P}_- \, \widetilde{D}_- \; = \; 0 \; ,
\\
&& \widetilde{P}_\pm \; = \; \frac{\mathds{1} \pm \gamma_5}{2} \,
\pm \, \widetilde{D}_- \, B^{-1} \, \frac{\mathds{1} - \gamma_5}{2}
\; , \label{ptilde}
\\
&& \widetilde{P}_+ \, + \, \widetilde{P}_- \, = \, \mathds{1} \quad
, \quad \widetilde{P}_\pm^2 \, = \, \widetilde{P}_\pm \quad , \quad
\widetilde{P}_+ \, \widetilde{P}_- \, = \, \widetilde{P}_- \,
\widetilde{P}_+ \, = \, 0 \; . \qquad
\end{eqnarray}
Comparing (\ref{ptilde}) with (\ref{latpplus}) and (\ref{latpminus})
one observes that the projectors acting from the left have the same
structure as the ones acting from the right, i.e., the terms
quadratic in $D_-,\widetilde{D}_-$ appear symmetrically.

\section{Blocking parameterized free fermions from the continuum}

In the last section we have discussed how the symmetries of left-handed 
continuum fermions are transported onto the lattice with the help of a
block-spin transformation. While above we considered the
blocking transformation in the background of a (suitably blocked) gauge field, 
we now discuss the free case where the blocking transformation from the
continuum can be evaluated in closed from. To obtain more general results we
use a parameterized continuum action where, depending on the values of 
some parameters, we interpolate between a chiral- and a vector-like theory. 
Furthermore we consider a parameterized blocking kernel which allows one
to turn on and off mixing between different components.  

\subsection{Blocked action for free parameterized fermions}

The parameterized continuum action we use as a starting point has the form

\begin{equation} S_{cont}[\overline{\Phi},\Phi] \; = \;
\int d^4 x \; \overline{\Phi}(x) \;
\overline{M}_{\overline{q}} \; \gamma_{\mu} \partial_{\mu} \; M_q \;
\Phi(x) \; , 
\end{equation}
where $ M_q, \overline{M}_{\overline{q}}$ denote matrices
in Dirac space, depending on a set of parameters $q, \overline{q}$.
A possible choice for such a matrix is 
\begin{equation}
M_q \; = \; q \, \frac{\mathds{1} + \gamma_5}{2} \, + \, \frac{\mathds{1} -
  \gamma_5}{2} \; .
\end{equation}
The real parameter $q$ allows to interpolate between
chiral ($q=0$) and vector-like theories ($q=1$). 
For $q \neq 0 \; M $ is invertible,
while for $q=0$, $M$ turns into a left-handed projector $P_-$ which has no inverse.

As in (\ref{effact}) we perform a block-spin transformation to obtain the
lattice action $S_{latt}[\overline{\psi},\psi]$,
\begin{equation}
e^{-\,S_{latt}[\overline{\psi},\psi]} \; = \; \int D[\overline{\Phi},\Phi] \;
e^{-\,S_{cont}[\overline{\Phi},\Phi] \; - \; [\overline{\psi} -
\overline{\Phi}^B] \; B_{r,s,t} \; [\psi - \Phi^B]} \;. 
\label{blockspin}
\end{equation}
In our blocking prescription also the blocking kernel $B_{r,s,t}$ is augmented
with real parameters $r$, $s$ and $t$. A possible choice is
\begin{equation}
B_{r,s,t} \; = \; \Big[ \, r\, \mathds{1} \, + \, s \, \gamma_4 \, \Big] \, 
\left[ \, t \, \frac{\mathds{1} + \gamma_5}{2} \, + \, \frac{\mathds{1} -
  \gamma_5}{2} \right ] \; .
\end{equation}
This rather general form allows to switch between several cases: For $r=1, \,
s=0$ one has maximal mixing between left- and right-handed components, while
for $r=0,\, s=1$ there is no mixing. In addition the second factor allows to
turn off ($t=0$) the right-handed components also in the blocking kernel. This
implies, that also the blocking kernel is not invertible for all values of the
parameters. 

Following \cite{wiese}, the blocked fields
$\overline{\Phi}^B, \Phi^B$ are constructed by integrating the
continuum fields over hypercubes $c_n$ centered at the points $n$ of
a four-dimensional lattice:
\begin{equation}
\overline{\Phi}_n^B \; = \; \int_{c_n} d^4 x \; \overline{\Phi}(x),
\qquad \Phi_n^B \; = \; \int_{c_n} d^4 x \; \Phi(x) \;.
\end{equation}
When switching to momentum space, the relation between the blocked and the
original continuum fields assumes the form (again we use lattice units,
i.e., $a=1$)
\begin{equation}
\hat{\Phi}^B (q) = \sum_{k \in \mathds{Z}^4} \hat{\Phi}(q + 2 \pi k)
\; \Pi(q + 2\pi k), \quad \Pi(q) = \prod_{\mu=1}^{4} \frac{2
\sin(q_{\mu}/2)}{q_{\mu}} \;.
\end{equation}
Working in momentum space, the block-spin transformation
(\ref{blockspin}) can be solved in closed form. Applying the techniques of
\cite{wiese} we end up with the
general result for the parameterized lattice action:
\begin{equation}
S_{latt}[\overline{\psi},\psi] \; = \; \frac{1}{(2\pi)^4} \int_{-\pi}^{\pi} d^4 q \,
\hat{\overline{\psi}}(q) \, \hat{D}(q) \, \hat{\psi} (-q) \; .
\end{equation}
\noindent The lattice Dirac operator is given by
\begin{equation}
\hat{D}(q) \; = \; \overline{M}_{\overline{q}} \; [ \; M_q \,
B^{-1}_{r,s,t} \, \overline{M}_{\overline{q}} \; - \; i \, \gamma_{\mu} \;
V_{\mu}(q) \; ]^{-1} \; M_q \; , 
\label{lattdirac}
\end{equation}
\noindent where
\begin{equation}
V_{\mu}(q) \; = \;  \sum_{k_{\mu} \in \mathbb{Z}^4} \, \Pi(q+2\pi
k)^2 \, \frac{(q+2\pi k)_{\mu}}{(q+2\pi k)^2} \; . 
\label{Dirac}
\end{equation}

Before discussing left-handed fermions in the next subsection, 
we stress that the parameterized formula was tested by setting $M_q =
\overline{M}_q = \mathds{1}$ and $B_{r,s,t} = 2 \mathds{1}$, which reproduces the 
vector-like lattice fermions of \cite{wiese}.

\subsection{Different choices of the parameters}

The result (\ref{lattdirac}), (\ref{Dirac}) with the free parameters $q,\,
\overline{q},\, r,\, s$ and $t$ provides an interesting laboratory for
exploring the interplay of the symmetries of the continuum action and the 
various choices of the blocking kernel.  

First we remark, that the result (\ref{lattdirac}) can be recast into a second
form which sheds a different light on the interplay between $M_q$,
$\overline{M}_{\overline{q}}$ and $B_{r,s,t}$. Furthermore in the alternative
form the inverse of the blocking kernel is not needed and it is
straightforward to perform a limit of the parameters where $B_{r,s,t}$ is
proportional to a projector. Simple algebraic manipulations
lead to 
\begin{equation}
\hat{D}(q) \; = \; \Big[ \;
\mathds{1} \,  \; - \; i \, B_{r,s,t} \; \gamma_{\mu} \;
M_q^{-1} \, V_{\mu}(q) \, \overline{M}_{\overline{q}}^{-1} \; \Big]^{-1} \; 
B_{r,s,t}\; . 
\label{lattdirac2}
\end{equation}

For the interpretation of the final results it is interesting to know the 
behavior of the functions $V_{\mu}(q)$. One can show that they  
behave as
\begin{equation}
V_{\mu}(q) \; \sim \; 
\left\{
  \begin{array}{cl}
    q_\mu/q^2 & \quad \mbox{for the physical branch} \, , \\
    0 & \quad \mbox{for the doublers} \; . \\
  \end{array}
\right.
\end{equation}

A possible naive choice which is expected to give free left-handed lattice
fermions is to set 
\begin{equation}
M_q \; = \; \overline{M}_{\overline{q}} \; = \; \mathds{1} \quad
, \quad B_{r,s,t} \; = \; c\, \gamma_4 \, \frac{\mathds{1} - \gamma_5}{2} \; .
\end{equation}
Thus in the blocking kernel we only allow left-handed components to survive. 
This choice leads to an operator of the form
\begin{equation}
D_- \; = \; \frac{(c+v_4) \ \gamma_4 + v_j \ \gamma_j}{(c+v_4)^2 + \vec{v}^2} \,
\frac{\mathds{1} - \gamma_5}{2} \; .
\end{equation}

This operator is free of doublers and reaches the correct limit of a single
left-handed fermion. However, a detailed analysis reveals 
singularities in momentum space for arbitrary values of $c$ which give
rise to a non-local action. 

An alternative choice which we currently explore, is to use a blocking 
kernel obeying
\begin{equation}
M_q \, B_{r,s,t}^{-1} \, \overline{M}_{\overline{q}} \; = \; Q \; ,
\label{blockansatz}
\end{equation}
with $Q$ being an arbitrary constant matrix. The target Dirac-operator has
the block-form
\begin{equation}
\left(
  \begin{array}{cc}
    0 & X \\
    0 & Y \\
  \end{array}
\right) 
\end{equation} 
with $2 \times 2$-matrices $X$ and $Y$.
By specifying the entries of $Q$, both $\chi$-GW equations might be fulfilled. 
The properties of this construction are currently explored.

In the last subsection we have stressed, that for some parameter values our
matrices $M_q$, $\overline{M}_{\overline{q}}$ and $B_{r,s,t}$ are not
invertible - in particular when they become projectors. However, the final
results (\ref{lattdirac}) and (\ref{lattdirac2}), respectively, 
contain products of these matrices (see e.g.~(\ref{lattdirac}))
such that singularities may cancel, and the limits where individual matrices
are singular might still lead to a finite answer. The ansatz
(\ref{blockansatz}) explores this possibility, but the consequences of this idea 
still have to be understood in detail. 

\section{Concluding remarks}

In this contribution we have explored the lattice representation of the
symmetries for left-handed continuum fermions. The tool we use is a
generating functional for the lattice theory obtained via blocking
from the continuum. With this technique we analyze how a symmetry of
the continuum theory manifests itself on the lattice. As a side
result we find that using the generating functional allows one also
to match the fermionic continuum and lattice measures and in
particular to map the anomalies.

We apply the approach to transferring the symmetry properties of
left-handed continuum fermions onto the lattice. It has to be
stressed that we do not map the chiral symmetry of a vector-like
theory, but instead directly transport the symmetries (\ref{symm1})
and (\ref{symm2}) onto the lattice, which are specific for
left-handed continuum fermions. This gives rise to the two $\chi$-GW
equations (\ref{latsymm1}) and (\ref{latsymm2}). These two equations
are the direct lattice manifestations of the continuum symmetries
for left-handed fermions without the intermediate step of using a
vector-like lattice theory. We show that a joint solution of the two
$\chi$-GW relations gives rise to a left-handed lattice Dirac
operator which also obeys the vector-like GW equation
(\ref{chirgw}), the lattice counterpart of the anti-commutator
(\ref{contcommut}) for left-handed continuum fermions. Furthermore we show
that a joint solution of the two $\chi$-GW generates a consistent algebra of
projectors. 

To analyze the approach further, and as a possible step towards an explicit
construction, we consider the blocking of free fermions. Both, the action and
the blocking kernel have free parameters which allow one to interpolate
between chiral and vector-like theories and to select how various components
couple in the blocking procedure. The blocked lattice action is given in
closed form, and currently different limits of the parameters, which could
give rise to chiral fermions are being studied.


\begin{thebibliography}{1234567}

\bibitem{GiWi82}
  P.H.~Ginsparg and K.G.~Wilson, Phys.\ Rev.\ D 25 (1982) 2649.

\bibitem{luscher}
  M.~L\"uscher,
  Phys.\ Lett.\  B 428 (1998) 342
  [arXiv:hep-lat/9802011].

\bibitem{hasenfratz1}
  P.~Hasenfratz, V.~Laliena and F.~Niedermayer,
  Phys.\ Lett.\  B 427 (1998) 125
  [arXiv:hep-lat/9801021].


\bibitem{overlap}
  R.~Narayanan and H.~Neuberger,
  Nucl.\ Phys.\  B 443 (1995) 305;
%
  H.~Neuberger,
  Phys.\ Lett.\  B 417 (1998) 141.

\bibitem{fp}
  P.~Hasenfratz and F.~Niedermayer,
  Nucl.\ Phys.\  B 414 (1994) 785
  [arXiv:hep-lat/9308004];
%
  P.~Hasenfratz, S.~Hauswirth, K.~Holland, T.~J\"org, F.~Niedermayer and U.~Wenger,
  Int.\ J.\ Mod.\ Phys.\  C 12 (2001) 691
  [arXiv:hep-lat/0003013].

\bibitem{chiralgauge} 
  M.~L\"uscher,
  Nucl.\ Phys.\  B 549 (1999) 295
  [arXiv:hep-lat/9811032];
%
  Nucl.\ Phys.\  B 568 (2000) 162
  [arXiv:hep-lat/9904009];
%
  JHEP 0006 (2000) 028
  [arXiv:hep-lat/0006014];
%
  H.~Neuberger,
  Phys.\ Rev.\  D 63 (2001) 014503
  [arXiv:hep-lat/0002032];
%
  D.~Kadoh and Y.~Kikukawa,
  arXiv:0709.3658 [hep-lat];
%
  arXiv:0709.3656 [hep-lat].

\bibitem{wiese}
  W.~Bietenholz and U.~J.~Wiese,
  Nucl.\ Phys.\  B 464 (1996) 319
  [arXiv:hep-lat/9510026];
  W.~Bietenholz and U.~J.~Wiese,
  Phys.\ Lett.\  B 378 (1996) 222
  [arXiv:hep-lat/9503022].

\bibitem{hasenfratz}
  P.~Hasenfratz, F.~Niedermayer and R.~von Allmen,
  JHEP 0610 (2006) 010
  [arXiv:hep-lat/0606021].

\bibitem{Fujikawa}
  K.~Fujikawa, Phys.~Rev.~Lett.~42 (1979) 1195; Phys.~Rev.~D 21 (1980) 2848.

\bibitem{NiNo}
  H.B.~Nielsen and M.~Ninomiya, Phys.\ Lett.\ B 105 (1981) 219.

\end{thebibliography}
\end{document}